\documentclass[10pt,english,two column]{IEEEtran}
\usepackage[T1]{fontenc}
\usepackage[latin9]{inputenc}
\usepackage{amsmath}
\usepackage{amssymb}
\usepackage{graphicx}
\usepackage{esint}

\makeatletter
\normalsize

\usepackage{units}

\ifCLASSINFOpdf
\else
\fi

\usepackage{babel}

\usepackage{babel}

\usepackage{babel}
\newcounter{mytempeqncnt}

\makeatother

\usepackage{babel}
\begin{document}

\title{Full-Duplex Cloud Radio Access Networks: An Information-Theoretic
Viewpoint}

\author{Osvaldo Simeone,%
\thanks{O. Simeone is with the ECE Dept, NJIT, Newark, USA. E-mail: osvaldo.simeone@njit.edu.%
}~ Elza Erkip,%
\thanks{E. Erkip is with the ECE Dept, NYU Polytechnic School of Engineering,
Brooklyn, NY, USA. E-mail: elza@poly.edu.%
}~ and Shlomo Shamai (Shitz)%
\thanks{S. Shamai is with the EE Dept, Technion, Haifa, Israel. E-mail: sshlomo@ee.technion.ac.il%
}}
\maketitle
\begin{abstract}
The conventional design of cellular systems prescribes the separation
of uplink and downlink transmissions via time-division or frequency-division
duplex. Recent advances in analog and digital domain self-interference
interference cancellation challenge the need for this arrangement
and open up the possibility to operate base stations, especially low-power
ones, in a full-duplex mode. As a means to cope with the resulting
\emph{downlink-to-uplink interference} among base stations, this letter
investigates the impact of the Cloud Radio Access Network (C-RAN) architecture.
The analysis follows an information theoretic approach based on the
classical Wyner model. The analytical results herein confirm the significant
potential advantages of the C-RAN architecture in the presence of full-duplex
base stations, as long as sufficient fronthaul capacity is available
and appropriate mobile station scheduling, or successive interference
cancellation at the mobile stations, is implemented.

\end{abstract}
\IEEEpeerreviewmaketitle

\section{Introduction}

The conventional design of cellular systems prescribes the separation
of uplink and downlink transmission via time-division or frequency-division
duplex. One of the main reasons for this choice is that operating
a base station in both the uplink and the downlink at the same time
causes the downlink transmitted signal to interfere with the uplink
received signal. This self-interference, if not cancelled, overwhelms
the uplink signal and makes the full-duplex operation of the base
station impractical. Recent advances in analog and digital domain
self-interference cancellation challenge the need for
this arrangement and open up the possibility to operate base stations,
especially low-power ones, in a full-duplex mode (see the review in
\cite{JSAC fd review}).

\begin{figure}[htbp!]
\centering \includegraphics[clip,width=2.3in]{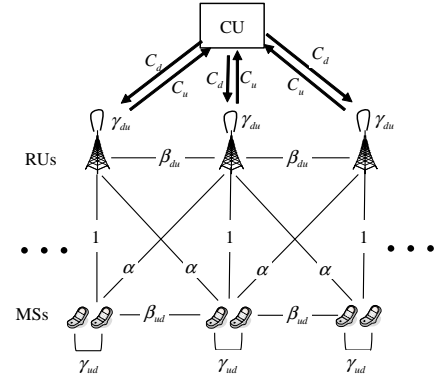} \caption{A full-duplex C-RAN based on a generalization of the Wyner model.}

\label{fig:1}
\end{figure}

Full-duplex base stations with effective self-interference cancellation
seemingly enable the throughput of a cellular system to be doubled,
since the available bandwidth can be shared by the uplink and the
downlink. However, this conclusion neglects two additional sources
of interference between uplink and downlink transmissions, namely:
(\emph{i}) the \emph{downlink-to-uplink (D-U) inter-cell interference}
that is caused on the uplink signals by the downlink transmissions
of neighboring base stations; and (\emph{ii}) the \emph{uplink-to-downlink
(U-D) interference }that is caused on the downlink signals by the
transmission of mobile stations (MSs), both within the same cell and
in other cells \cite{Choi STD}. The impact of intra-cell U-D interference
has been studied in \cite{Goyal} and \cite{Barghi} for a single-cell
system with a single-antenna or a multi-antenna base station, respectively.
Multi-cell systems, in which D-U interference and also inter-cell
U-D interference arise, have been studied in \cite{Choi STD} via
system simulation and in \cite{Goyal0} using stochastic geometry.

The prior work mentioned above focuses on single-cell processing techniques,
in which baseband processing is carried out locally at the base stations.
Single-cell processing is inherently limited by the D-U interference.
With the aim of overcoming this limitation, here we investigate the
impact of the Cloud Radio Access Network (C-RAN) architecture on a
full-duplex cellular system. In a C-RAN system, the base stations operate
solely as radio units (RUs), while the baseband processing is carried
out at a central unit (CU) within the operator's network \cite{China}.
This migration of baseband processing is enabled by a network of fronthaul
links, such as fiber optics cables or mmwave radio links, that connect
each RU to the CU. The centralization of both uplink and downlink
baseband processing at the CU allows the CU to perform cancellation
of the D-U interference since the downlink signal is known at the
CU. In order to further cope also with the U-D interference, we evaluate
the advantages of performing successive interference cancellation
at the MSs. Accordingly, the strongest intra-cell uplink transmissions
are decoded and cancelled before decoding the downlink signals.

The analysis in this letter takes an information theoretic approach
that builds on the prior work reviewed in \cite{Simeone:FnT}. Specifically,
in order to capture the key elements of the problem at hand,
with particular emphasis on the various sources of interference, we
focus on a modification of the classical Wyner model. The adoption of this model enables us to derive
analytical expressions for the achievable rates under single-cell
processing and C-RAN operation assuming either half-duplex or full-duplex
base stations. These analytical results provide obtain fundamental insights
into the regimes in which full-duplex RUs, particularly when implemented
with a C-RAN architecture, are expected to be advantageous.

\section{System Model and Notation}

Consider the extended Wyner model depicted in Fig. \ref{fig:1}. The
model contains one MS per cell that transmits in the uplink and one
that receives in the downlink. With conventional half-duplex RUs,
the two MSs transmit in different time-frequency resources, while,
with full-duplex RUs, uplink and downlink are active at the same time.
We describe here the system model for the full-duplex system -- the
modifications needed to describe the half-duplex system will be apparent.

There are $N$ cells and inter-cell interference takes place only
between adjacent cells as shown in Fig. \ref{fig:1}. In order to
avoid border effects, as it is customary, we take $N$ to be very
large. Due to the limited span of the interference, results in the
regime of $N\rightarrow\infty$ are known to be accurate also for
small values of $N$ (see \cite{Simeone:FnT}). In the uplink, the
MS active in the $k$th cell transmits a signal $x_{u,k}$ with power
$p_{u}=E[|x_{u,k}|^{2}]\leq P_{u}$, with $P_{u}$ being the power
constraint. Similarly, in the downlink, each $k$th RU transmits a
signal $x_{d,k}$ with power $p_{d}=E[|x_{d,k}|^{2}]\leq P_{d}$.
The baseband signal received in uplink by the $k$th RU is given as

\begin{equation}
y_{u,k}=h_{k}\star x_{u,k}+h_{du,k}\star x{}_{d,k}+z_{d,k},\label{eq:uplink model}
\end{equation}
where $\star$ denotes the convolution; $h_{k}=\delta_{k}+\alpha\delta_{k-1}+\alpha\delta_{k+1}$,
where $\delta_{k}$ is the Kronecker delta function, accounts for
the direct channel, which has unit power gain, and for the inter-cell
interference, which is characterized by the inter-cell interference
power gain $\alpha^{2}$; $h_{du,k}=\beta_{du}\delta_{k-1}+\gamma_{du}\delta_{k}+\beta_{du}\delta_{k+1}$
models the\emph{ D-U interference} with inter-cell power gain $\beta_{du}^{2}$
and self-interference power gain $\gamma_{du}^{2}$; and $z_{u,k}$
is white Gaussian noise with unit power.

In the downlink, the signal received by the MS in the $k$th cell
can be written as
\begin{equation}
y_{d,k}=h_{k}\star x_{d,k}+h_{ud,k}\star x_{u,k}+z_{d,k},\label{eq:downlink model}
\end{equation}
where $h_{ud,k}=\beta_{ud}\delta_{k-1}+\gamma_{ud}\delta_{k-1}+\beta_{ud}\delta_{k+1}$
describes the \emph{U-D interference}, which has inter-cell power
gain $\beta_{ud}^{2}$ and intra-cell power gain $\gamma_{ud}^{2}$;
and $z_{d,k}$ is white Gaussian noise with unit power. As depicted
in Fig. \ref{fig:1}, the parameter $\gamma_{ud}^{2}$ accounts for
the power received by the MS active in the downlink from the MS active
in the uplink within the same cell.

Each RU is connected to the CU with a fronthaul link of capacity $C_{u}$
in the uplink and $C_{d}$ in the downlink. These capacities are measured
in bits/s/Hz, where the normalization is with respect to the bandwidth
shared by the uplink and downlink channels.

We assume full channel state information at the CU for both uplink
and downlink. Define as $R_{d}$ and $R_{u}$ the per-cell rates,
measured in bits/s/Hz, achievable in uplink and downlink, respectively,
by a particular scheme. The \emph{equal per-cell rate} is now defined
as $R_{eq}=\min\left\{ R_{u},R_{d}\right\} .$

\emph{Notation}: For convenience of notation, we define the Shannon
capacity $\mathrm{C}(S)=\log_{2}(1+S)$ and the function $\mathrm{q}(a,b,c)=\textrm{min\ensuremath{(a,}max\ensuremath{(b,c))}}$.

\section{Half-Duplex Operation}

In this section, we review the performance in the presence of the
conventional half-duplex constraint on the RUs. In this case, a fraction
$f\in[0,1]$ of the time-frequency resources are devoted to the uplink
and the remaining fraction $1-f$ to the downlink.

\subsection{Single-Cell Processing}

With single-cell processing, each RU encodes in downlink and decodes
in uplink with no cooperation from the other RUs. The fronthaul links
are used to convey the downlink information streams from the CU to
the RUs and to transport the decoded uplink data streams from the
RUs to the CU.

In the uplink, the inter-cell interference, which has power $2\alpha^{2}P_{u}$,
is treated as noise. As a result, the achievable rate per cell is
given as
\begin{equation}
R_{u}=\min\left\{ \mathrm{C}\left(P_{u}/(1+2\alpha^{2}P_{u})\right),C_{u}\right\} .\label{eq:uplink HD SCP}
\end{equation}
In (\ref{eq:uplink HD SCP}), the second term accounts for the limitations
imposed by the fronthaul links for the transmission of the decoded
data streams to the CU. Similarly, by treating inter-cell interference
as noise in the downlink, we obtain the per-cell achievable rate $R_{u}=\min\left\{ \mathrm{C}\left(P_{d}/(1+2\alpha^{2}P_{d})\right),C_{d}\right\} $.
We note that full power is used in both uplink and downlink with no
loss of optimality. Finally, the equal per-cell rate is obtained by
optimizing over the fraction $f$ as $R_{eq}=\underset{f\in[0,1]}{\textrm{max}}\min\left\{ fR_{u},(1-f)R_{d}\right\} $,
which yields
\begin{align}
R_{eq} & =R_{d}R_{u}/(R_{d}+R_{u}).\label{eq:Req}
\end{align}

\subsection{C-RAN}

With C-RAN operation, baseband processing is carried out at the CU,
while the RUs act solely as downconverters in the uplink and upconverters
in the downlink. Unlike the single-cell processing case, the fronthaul
links here carry compressed baseband information.

\subsubsection{Uplink\label{sub:Uplink}}

In the uplink, the signals received by the RUs are compressed and
forwarded to the CU, which then performs joint decoding. To elaborate,
each $i$th RU produces the compressed version of the received signal
\begin{equation}
\hat{y}_{u,i}=y_{u,i}+q_{u,i},\label{eq:quantized uplink signal}
\end{equation}
where $q_{u,i}\sim\mathcal{CN}(0,\sigma_{u}^{2})$ is the quantization
noise, which is white and independent of all other variables. Using
standard results in rate-distortion theory (see, e.g., \cite[Sec. 3.6]{ElGamal}),
assuming separate decompression at the CU for each fronthaul link,
the quantization noise power is obtained by imposing the equality
$C_{u}=I(y_{u,i};\hat{y}_{u,i})$, which yields
\begin{equation}
\sigma_{u}^{2}=(1+(1+2\alpha^{2})P_{u})/(2^{C_{u}}-1).\label{eq:quant uplink}
\end{equation}
Based the received signals (\ref{eq:quantized uplink signal}), the
CU performs joint decoding. The corresponding achievable rate per
cell can be written as $R_{u}=\lim_{N\rightarrow\infty}I(\mathbf{x}_{u}^{N};\mathbf{\hat{y}}_{u}^{N})/N$,
where we have defined $\mathbf{x}_{u}^{N}=[x_{u,1}\cdots x_{u,1N}]^{T}$
and similarly for $\mathbf{\hat{y}}_{u}^{N}$. The limit at hand can
be calculated as (see, e.g., \cite[Sec. 3.1.2]{Simeone:FnT})
\begin{equation}
R_{u}=\int_{0}^{1}\mathrm{C}\left(P_{u}H(f){}^{2}/(1+\sigma_{u}^{2})\right)df,\label{eq:uplink HD CRAN}
\end{equation}
where $H(f)=1+2\alpha\cos(2\pi f)$ is the Fourier transform of $h_{k}$.
Note that in (\ref{eq:uplink HD CRAN}) the quantization noise affects
the noise level at the decoder.

\subsubsection{Downlink\label{sub:Downlink HD}}

For the downlink, the CU performs channel coding and precoding and
then compresses the resulting baseband signals prior to transmission
on the fronthaul links to the RUs. The RUs then simply upconverts
the baseband signal and transmits it to the MSs. We assume that channel
encoding is performed separately on each data stream producing the
independent signal $s_{k}\sim\mathcal{CN}(0,P_{s})$ for each $k$th
cell. The selection of the power $P_{s}$ will be discussed below.
Linear precoding is then applied at the CU, so that the precoded signal
reads $\tilde{x}_{d,k}=g_{k}\star s_{k}$ for a given precoding vector
$\mathbf{g}=\{g_{k}\}_{k=-\infty}^{+\infty}$. Without loss of generality,
we impose the constraint $||\mathbf{g}||^{2}=1$. One can also assume
that, by the symmetry of the problem, the filter $\mathbf{g}$ is
real and symmetric around $k=0$, i.e., $g_{k}=g_{-k}$. The precoded
signal is quantized producing the quantized baseband signal
\begin{equation}
x_{d,k}=\tilde{x}_{d,k}+q_{d,k}\label{eq:quant noise downlink}
\end{equation}
with quantization noise $q_{d,i}\sim\mathcal{CN}(0,\sigma_{d}^{2})$,
which is white and independent of all other variables. The quantization
noise is related to the fronthaul capacity $C_{d}$ by imposing the
equality $C_{d}=I(\tilde{x}_{d,k};x_{d,k})$, which yields $\sigma_{d}^{2}=P_{s}/(2^{C_{d}}-1).$

The achievable per-cell rate is
\begin{equation}
R_{d}=\mathrm{C}\left(\frac{P_{s}\tilde{h}_{0}^{2}}{1+2P_{s}\sum_{k>0}\tilde{h}_{k}^{2}+\sigma_{d}^{2}(1+2\alpha^{2})}\right).\label{eq:downlink HD CRAN}
\end{equation}
with $\tilde{h}_{k}=h_{k}\star g_{k}$. Note that in (\ref{eq:downlink HD CRAN})
the sources of noise are the interference from the undesired downlink
signal streams ($2P_{s}\sum_{k>0}\tilde{h}_{k}^{2}$) and the quantization
noise ($\sigma_{d}^{2}(1+2\alpha^{2})$). The power $P_{s}$ is obtained
by enforcing the power constraint, namely $P_{d}=E[|x_{d,k}|^{2}]=P_{s}+\sigma_{d}^{2}$,
which leads to $P_{s}=P_{d}(1-2^{-C_{d}}).$ We observe that, in the
special case in which zero-forcing (ZF) linear precoding is adopted,
we have $\tilde{h}_{k}^{2}=0$ for $k\neq0$ and $\tilde{h}_{0}^{2}=(\int_{0}^{1}H(f)^{-2}df)^{-1}=(1-4\alpha^{2})^{3/2}$
in (\ref{eq:downlink HD CRAN}) (see \cite[Sec. 4.2.3]{Simeone:FnT}). In summary,
for any given precoding filter $\mathbf{g}$, the per-cell equal rate
is equal to (\ref{eq:Req}) with $R_{u}$ in (\ref{eq:uplink HD CRAN})
and $R_{d}$ in (\ref{eq:downlink HD CRAN}).

\section{Full-Duplex Operation\label{sec:Full-Duplex-Operation}}

In this section, we consider the performance with full-duplex RU operation. As in \cite{Goyal}-\cite{Goyal0}, we assume that the cancellation of known D-U interference signals is ideal in order to focus on the potential advantages of full-duplex.

\subsection{Single-Cell Processing\label{sub:Single-Cell-Processing}}

With single-cell processing, each RU is able to cancel its self-interference D-U signal. As a result, the achievable uplink per-cell rate is
obtained, similar to (\ref{eq:uplink HD SCP}), as $R_{u}=\min\left\{ \mathrm{C}\left(p_{u}/(1+2\alpha^{2}p_{u}+2\beta_{du}^{2}p_{d})\right),C_{u}\right\} ,$
where the additional term $2\beta_{du}^{2}p_{d}$
at the denominator accounts for the D-U interference. Note that we
have allowed for a transmit power $p_{u}=E[|x_{u,k}|^{2}]\leq P_{u},$
since with full-duplex, unlike the case of half-duplex operation,
it can be advantageous not to use the full available power. In an
analogous fashion, the achievable rate for the downlink is $R_{d}=\min\left\{ \mathrm{C}\left(p_{d}/(1+2\alpha^{2}p_{d}+(2\beta_{ud}^{2}+\gamma_{ud}^{2})p_{u})\right),C_{d}\right\} ,$
where the additional term $(2\beta_{ud}^{2}+\gamma_{ud}^{2})p_{u}$
is the power of the U-D interference.

The intra-cell interference, with power $\gamma_{ud}^{2}p_{u}$, is
caused by a MS in the same cell and therefore it is expected to be
very relevant. In order to mitigate this problem, here we explore
the possibility that the MSs implement a successive interference cancellation
receiver in which the intra-cell uplink signal is first decoded and
then cancelled before decoding the intended signal. Now, using standard
results on the capacity region of multiple access channels (see, e.g.,
\cite[Sec. 4.6]{ElGamal}), the rate achievable with single-cell processing
is given as $R_{d}=\min\left\{ \mathrm{q}(t_{1},t_{2}-R_{u},t_{3}),C_{d}\right\} $
with $t_{1}$, $t_{2}$ and $t_{3}$ equal to $\mathrm{C}\left(p_{d}/(1+2\alpha^{2}p_{d}+2\beta_{ud}^{2}p_{u})\right)$,
$\mathrm{C}\left((p_{d}+\gamma_{ud}^{2}p_{u})/(1+2\alpha^{2}p_{d}+2\beta_{ud}^{2}p_{u})\right)$
and $\mathrm{C}\left(p_{d}/(1+2\alpha^{2}p_{d}+(2\beta_{ud}^{2}+\gamma_{ud}^{2})p_{u})\right)$, respectively.
Finally, the equal per-cell rate can be calculated as
\begin{equation}
R_{eq}=\underset{p_{u}\leq P_{u},\textrm{ }p_{d}\leq P_{d}}{\textrm{max}}\min\left\{ R_{u},R_{d}\right\} .
\end{equation}

% ensure that we have normalsize text
% Store the current equation number.
\begin{figure*}[!t]
\setcounter{mytempeqncnt}{\value{equation}} % Set the equation number to one less than the one
% desired for the first equation here.
% The value here will have to changed if equations
% are added or removed prior to the place these
% equations are referenced in the main text.
\setcounter{equation}{12}
\begin{equation}
R_{d}=\mathrm{C}\left(\frac{p_{d}(1-2^{-C_{d}})\tilde{h}_{0}^{2}}{1+2p_{d}(1-2^{-C_{d}})\sum_{k>0}\tilde{h}_{k}^{2}+(2\beta_{ud}^{2}+\gamma_{ud}^{2})p_{u}+p_{d}2^{-C_{d}}(1+2\alpha^{2})}\right).\label{eq:downlink FD CRAN}
\end{equation}
% Restore the current equation number.
\setcounter{equation}{\value{mytempeqncnt}} % IEEE uses as a separator
\hrulefill{}% The spacer can be tweaked to stop underfull vboxes.
\vspace*{4pt}
\end{figure*}

\subsection{C-RAN}

\subsubsection{Uplink}

As discussed above, in the uplink of a C-RAN, the signals received
by the RUs, after D-U self-interference cancellation, are compressed and forwarded to the CU, which then performs
joint decoding. Each $i$th RU produces a compressed version (\ref{eq:quantized uplink signal})
of the received signal. Similar to (\ref{eq:quant uplink}), the quantization
noise power $\sigma_{u}^{2}$ is calculated as
\begin{equation}
\frac{1+(1+2\alpha^{2})p_{u}+2\beta_{du}^{2}(1+R_{g}(2))p_{d}}{2^{C_{u}}-1},\label{eq:quant noise FD uplink}
\end{equation}
where $R_{g}(\tau)=\sum_{k}g_{k}g_{k-\tau}$ is the correlation function
of the downlink filter.
Note that the third term in the numerator of (\ref{eq:quant noise FD uplink})
quantifies the contribution of the D-U interference, which is given
as $E[|\beta_{du}x_{k-1}+\beta_{du}x_{k+1}|^{2}]$.
Based the received signals (\ref{eq:quantized uplink signal}), the
CU first cancels the D-U interference. Note that this is possible
since the downlink signals are known to the CU. Then, the CU performs
joint decoding. Similar to (\ref{eq:uplink HD CRAN}), the corresponding
achievable rate per cell can be written as
\begin{equation}
R_{u}=\int_{0}^{1}\mathrm{C}\left(p_{u}H(f){}^{2}/(1+\sigma_{u}^{2})\right)df.\label{eq:uplink FD CRAN}
\end{equation}

\subsubsection{Downlink}

We adopt linear precoding as discussed in Sec. \ref{sub:Downlink HD}.
Accordingly, if U-D intra-cell interference is treated as noise, the
achievable per-cell rate is given as (\ref{eq:downlink FD CRAN}).
If instead successive interference cancellation is performed at the
MSs, the rate achievable in C-RANs can be written as $R_{d}=\mathrm{q}(t_{1},t_{2}-R_{u},t_{3})$,
where $t_{1}$, $t_{2}$ and $t_{3}$ can be calculated similar to
Sec. \ref{sub:Single-Cell-Processing} from (\ref{eq:downlink FD CRAN}).
For instance, $t_{1}$ equals (\ref{eq:downlink FD CRAN}) but with
the term $\gamma_{ud}^{2}p_{u}$ removed from the denominator. In
summary, for any given precoding filter $\mathbf{g}$, the per-cell
equal rate is equal to (\ref{eq:Req}) with $R_{u}$ in (\ref{eq:uplink FD CRAN})
and $R_{d}$ in (\ref{eq:downlink FD CRAN}).

\begin{figure}[htbp!]
\centering \includegraphics[clip,width=2.6in]{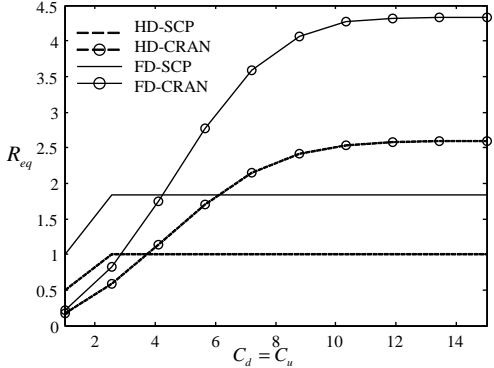} \caption{Equal per-cell rate $R_{eq}$ versus the fronthaul capacities $C_{d}=C_{u}$
with $P_{u}=P_{d}=20dB$ $\alpha=0.4$, $\beta_{du}=0.4$, $\beta_{ud}=0.04$,
and $\gamma_{ud}=4$. }

\label{fig:1-1}
\end{figure}

\section{Numerical Results and Concluding Remarks}

In this section, we provide some numerical results to bring insights into the performance of the discussed approaches. Fig. \ref{fig:1-1}, we plot the equal per-cell rate
versus the fronthaul capacities $C_{d}=C_{u}$ with $P_{u}=P_{d}=20dB$
$\alpha=0.4$, $\beta_{du}=0.4$, $\beta_{ud}=0.04$, $\gamma_{du}=0$
and $\gamma_{ud}=4$. Note that the parameter $\gamma_{du}$ does not play a role in the analysis. The inter-cell D-U interference
gain $\beta_{du}$ is chosen to be comparable to the inter-cell gain $\alpha$,
while the U-D intra-cell interference gain $\gamma_{ud}$ is significantly
larger and the corresponding inter-cell gain $\beta_{ud}$ is instead
significantly smaller than $\alpha$. This setting appears to be in
line with what is expected in a dense small-cell scenario in which
the RUs are placed in a more advantageous position than the MSs. A
ZF precoder is assumed for the downlink, and, unless stated otherwise,
successive interference cancellation (SIC) is employed in the downlink.

The figure shows that C-RAN solutions have a significant advantage
over the corresponding single-cell processing (SCP) approaches for
both half-duplex (HD) and full-duplex (FD) operations as long as the
fronthaul capacities are large enough. Note that the spectral efficiency
of the fronthaul links is expected to at least one order of magnitude
larger than the downlink or uplink spectral efficiencies, which is
well within the range shown in the figure. Moreover, when the fronthaul capacities are sufficiently large,
FD-C-RAN provides a gain of around 1.7 as compared to HD-C-RAN, which falls short of the maximum gain of 2 due to the interference between
uplink and downlink.

\begin{figure}[htbp!]
\centering \includegraphics[clip,width=2.6in]{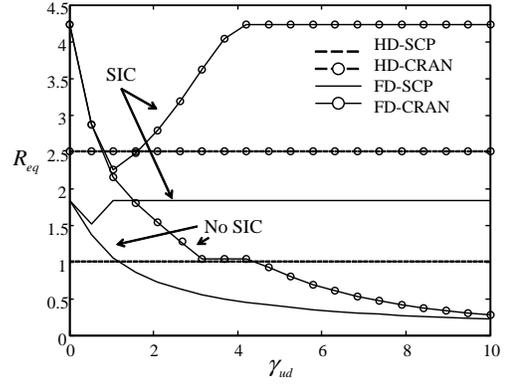} \caption{Equal per-cell rate $R_{eq}$ versus the U-D intra-cell interference
gain $\gamma_{ud}$ with $P_{u}=P_{d}=20dB$ $\alpha=0.4$, $\beta_{du}=0.4$,
$\beta_{ud}=0.04$, and $C_{u}=C_{d}=10$. }

\label{fig:1-3}
\end{figure}

We finally study the impact of U-D intra-cell interference in Fig. \ref{fig:1-3}. The parameters are the same as
for the previous figure. For the full-duplex approaches, we consider
the rate $R_{eq}$ achievable with and without SIC as a function of
$\gamma_{ud}$. It is seen that, FD-C-RAN is advantageous only if we
have small intra-cell interference $\gamma_{ud}$ or if the MSs implement
SIC and the gain $\gamma_{ud}$ is large enough. This suggests that,
in practice, FD-C-RAN should only be used in conjunction with an appropriate
scheduling algorithm that ensures one of these two conditions to be
satisfied.

Overall, the results herein confirm the significant potential advantages
of the C-RAN architecture in the presence of full-duplex base stations,
as long as sufficient fronthaul capacity is available and appropriate
MS scheduling or successive interference cancellation at the MSs is
implemented.

\end{document}